\documentclass[12pt]{article}
\usepackage{eurosym}
\usepackage{amssymb}
\usepackage{epsfig}
\usepackage{graphicx}
\usepackage{amsmath}
\usepackage{amsfonts}
\usepackage{amssymb}
\usepackage{float}
\usepackage{subfigure}
\usepackage{float}
\usepackage{amsmath}
\usepackage{color}
\usepackage{amsfonts}
\usepackage{amssymb}
\DeclareGraphicsExtensions{.pdf}
\graphicspath{{Images/}}
\begin{document}
\date{}
\title{\textbf{\Large Holographic dark energy in Barrow cosmology with Granda-Oliveros IR cutoff}}
\author{ \textbf{{\normalsize M. Motaghi$^{1}$}}, \textbf{{\normalsize A. Sheykhi$^{1,2}$}\thanks{%
            asheykhi@shirazu.ac.ir}}, \textbf{{\normalsize E. Ebrahimi$^{1,2}$}\thanks{%
es.ebrahimi@saadi.shirazu.ac.ir}} \\
$^{1}$ {\normalsize Department of Physics, College of Science,}
{\normalsize Shiraz University, Shiraz 71454, Iran}\\
$^{2}$ {\normalsize Biruni Observatory, College of Science, Shiraz
University, Shiraz 71454, Iran}}
\date{}
\maketitle

\begin{abstract}
Applying the modified Barrow entropy, inspired by the quantum
fluctuation effects, to the cosmological background, and using
thermodynamics-gravity conjuncture, the Friedmann equations get
modified as well. In this paper, we explore the holographic dark
energy with Granda-Oliveros (GO) IR cutoff, in the context of the
modified Barrow cosmology. First, we assume two dark components of
the universe evolves independently and obtain the cosmological
parameters and explore the cosmic evolution. Second, we consider
an interaction term between dark energy (DE) and dark matter (DM).
We observe that the Barrow parameter $\delta$ crucially affects
the cosmic dynamics, causes the transition from the decelerating
phase to the accelerating phase occurs later. We find out that the
equation of state parameter is in the quintessence region in the
past and crosses the phantom divide at the present time. Finally,
we examine the squared speed of sound analysis for this model.
According to the squared sound speed diagrams, the results
indicate that the presence of interaction between DM and DE  as
well as increasing in the value of $\delta$ leads to the
manifestation of signs of instability in the past $(v_s^2<0)$.
Furthermore, by examining the statefinder, we find that presence
of  $\delta$ also makes a distinction between holographic dark
energy in Barrow cosmology with GO-IR cutoff and the $\Lambda$CDM
model. In fact, increasing $\delta$ causes the statefinder diagram
move away from the point of $\left\lbrace r,s\right\rbrace=
\left\lbrace 1,0\right\rbrace$ at $z=0$.
\end{abstract}
\vspace*{0.2cm}
\section{Introduction}\label{1}
Almost around the end of last century, two groups of researchers
named the ``\textit{High-z Supernova Search Team}" and the
``\textit{Supernova Cosmology Project}", by analyzing the data
from high redshift type Ia supernovaes, independently discovered
that our Universe is currently undergoing a phase of accelerated
expansion \cite{riess, perlmutter}. This discovery was so
important that it shook the foundations of modern cosmology and
changed our view of the universe.  In the context of the standard
cosmology, the component of energy which is responsible for
accelerated expansion is called DE. What we know about DE, till
now, is that it has negative pressure with anti-gravity nature,
filled smoothly all space, and push our universe to accelerate.
The microscopic structure of the DE energy is still a mystery in
the modern cosmology.

Nowadays, the DE has been widely accepted as a new component of
energy in our Universe which is not only responsible for the
accelerated expansion, but its presence is necessary for
consistency of other observational data in the standard cosmology,
such as cosmic microwave background (CMB) anisotropy, large scale
structure and the age of the universe. Among all candidate for DE,
the cosmological constant, $\Lambda$, is located at the center,
both from theoretical and observational evidences. Although the
$\Lambda$CDM model is highly consistent with observations, but it
suffers some problems such a fine tunning and coincidence
problems. Besides, the equation of state (EoS) parameter of
$\Lambda$ is fixed, namely $w_{\Lambda}=p/\rho=-1$, while some
cosmological observations indicate a time dependent EoS parameter
for DE. Furthermore, recent observations reveal discrepancies with
the $\Lambda$CDM which are known as cosmic tensions \cite{Lusso}.
These tensions could be summarized as $H_0$ \cite{divalentino H0}
and $\sigma_8$ \cite{Riess2019, Riess2021} tensions. Thus the
branch of dynamical DE is developed.

Among all candidates for DE, the so called holographic dark energy
(HDE) which is based on the holographic principle has got a lot of
attentions in the literatures. According to the holographic
principle, the information inside a system can be encoded on its
boundary. In the context of black hole physics the holographic
principle is much well-known. It has been shown that the entropy
of a black hole, which show the number of degrees of freedom of
the system, is scaled by its horizon area rather than its volume
\cite{Bousso,hooft, susskind}. In this regard, Cohen et al.,
proposed that in quantum field theory a short distance cutoff
corresponds to a long distance cutoff due to the limit set by the
formation of a black hole, i.e. if $\rho$ is the resulting quantum
zero-point energy density, with a short distance cutoff, the total
energy in a region of size $L$ must not exceed the mass of a black
hole of the same size, hence, $L^3 \rho \le L M^2_{p}$
\cite{Cohen}. The largest length, known as the infrared length,
saturates this inequality. Thus one can define the energy density
of DE based on the holographic principle as \cite{Wang}
\begin{align}\label{rho1}
    \rho_{de}=3c^2M_p^2L^{-2},
\end{align}
where $3c^2$ is a numerical constant given for the convenience of
calculations, $M_p^2=(8\pi G)^{-1}$ is the reduced Planck mass,
and $L$ is defined as the infrared (IR) cutoff and is related to
the size of the current universe. The simplest choice for $L$ is
the Hubble radius, $L=1/H$. In a universe filled by a DE component
an accelerated expansion occurs when $w_{de}<-1/3$, while it was
pointed out that for $L=1/H$ the EoS parameter becomes zero
($w_{de}=0$) \cite{Hsu}. If one choose the particle horizon as
infrared cutoff, one arrives at $w_{de}>-1/3$, while in the case
of the future event horizon as IR cutoff, the condition
$w_{de}<-1/3$ is satisfied \cite{Li}. The latter suffers the
causality problem \cite{Li}. Thus a plenty models of HDE are
introduced in the literature. For a detailed review on the HDE
models one can see \cite{lireview} and references therein. A
special choice for IR cutoff was proposed by Granda and Oliveros
\cite{Granda1,Oliveros}, which in addition to $H$, it includes
time derivative of the Hubble parameter, and is written as
\begin{equation}\label{lgo}
L=(\alpha H^{2}+\beta \dot{H})^{-1/2}.
\end{equation}
where $\alpha$ and $\beta$ are arbitrary constants.

Inspired by the structure of the COVID-19 virus, J. D. Barrow
\cite{Barrow} argued that the quantum gravitational effects could
cause changes in the geometry of the black hole horizon. The
change in the geometrical structure affects the form of the
horizon area and consequently the corresponding black hole entropy
get modified as \cite{Barrow}
\begin{align}\label{bentropy}
    S=\left(\dfrac{A}{A_0} \right)^{1+\delta/2},
\end{align}
where $A$ is the standard horizon area, $A_0$ the Planck area, and
$\delta$ represents the amount of quantum gravitational
deformation effects which ranges as $0<\delta<1$. Note that
$\delta=1$ corresponds to the most complex and fractal structure
of the black hole horizon, while for $\delta=0$ and $A_0=4G$, the
standard Bekenstein-Hawking entropy of the black hole is restored.
Since the HDE model is related to the geometry of the boundary,
any modification to the entropy causes a modification on the DE
density. Therefore, the Barrow holographic dark energy (BHDE)
density is defined as \cite{Saridakis}
\begin{align}\label{rh}
    \rho_{de}=C L^{\delta-2},
\end{align}
where $C$ is a parameter with dimensions
$[L]^{-2-\delta}$ . When $\delta=0$, Eq. \eqref{rho1}
is reproduced for $C=3c^2M_p^2$.

On the other hand, it is a general belief that there is a deep
connection between the gravitational field equations and the laws
of thermodynamics (see \cite{Jacobson}-\cite{Abreu} and references
therein). This connection has also been confirmed in the context
of cosmology, where it has been shown that the Friedmann equations
describing the evolution of the universe can be written as the
first law of thermodynamics on the apparent horizon and vice versa
\cite{Cai2}-\cite{Sheykhi4}. This derivation is crucially depends
on the form of the entropy expression. Any modification to the
entropy expression changes the corresponding Friedmann equations.
When the entropy of the horizon is in the form of Barrow entropy
(\ref{bentropy}), the modified Friedman equations were extracted
in \cite{Saridakis1,Sheykhi5}. Modified cosmology through Barrow
entropy were explored in \cite{Sheykhi6}. It is important to note
that the exponent $\delta$ in Barrow entropy, cannot reproduce any
term which may play the role of DE and one still needs to take
into account the DE (cosmological constant) component in the
Friedmann equations to reproduce the accelerated universe
\cite{Sheykhi6}. In \cite{Saridakis1}, the authors discussed that
the new extra terms that constitute an effective energy density
sector are appeared in the Friedmann equations based on Barrow
entropy. This indicates that Barrow cosmology provides a new
background for study various models of DE. The aim is to reveal
the influences of the Barrow exponent $\delta$ on the cosmological
parameters and phase transition of the universe. Based on the
modified Barrow entropy, a new HDE model has been proposed by
choosing various IR cutoff
\cite{Saridakis:2020zol,Srivastava:2020cyk,Adhikary:2021xym,Anagnostopoulos:2020ctz,Dabrowski:2020atl,Oliveros:2022biu,SheBHDE,SheBADE}.
Applying the spherical collapse formalism, growth of perturbation
in the context of Barrow cosmology were explored in
\cite{Sheykhi:2022gzb}.

Let us note that in most studies on HDE, based on Barrow entropy,
the authors only modify the energy density of HDE, while they
still use the Friedmann equations in standard cosmology
\cite{Saridakis:2020zol,Srivastava:2020cyk,Adhikary:2021xym,Anagnostopoulos:2020ctz,Dabrowski:2020atl,Oliveros:2022biu}.
This is inconsistent with the thermodynamics-gravity conjecture,
which implies that the Friedmann equations get modified due to
modification to the entropy expression. Thus, one should consider
the modified energy density of HDE in the context of modified
Barrow cosmology. This fact were already considered for studying
both HDE and ADE in the background of Barrow cosmology
\cite{SheBHDE,SheBADE}. In the present work we shall study Barrow
HDE (BHDE) in the background of the modified Barrow cosmology. Our
work differs from \cite{Oliveros:2022biu} in that we consider BHDE
with GO cutoff in the background of modified Friedmann equations,
while the authors of \cite{Oliveros:2022biu} only modifies the
energy density, and explores its consequences in the context of
standard cosmology.

Our work is organized as follows. In section \ref{Ther}, we use
the correspondence between thermodynamics and gravity and extract
the modified Friedman equations based on Barrow entropy. In
section \ref{Model}, we study BHDE with GO IR cutoff in the
background of Barrow cosmology in a flat universe. In section
\ref{Nonflat}, we extend our study to the case of a non-flat
universe. In section \ref{Stab}, we explore stability and
statefinder of the model. The last section is devoted to
conclusions.
\section{Modified Friedman equations from Barrow entropy}\label{Ther}
We start by deriving the modified Friedmann equation inspired by
Barrow entropy. Using thermodynamics-gravity conjecture, the
cosmological field equations from modified Barrow entropy were
extracted in \cite{Sheykhi5}. The cosmological consequences of the
obtained modified Friedmann equations has been also explored
\cite{Sheykhi6}. In this section, following \cite{Sheykhi5}, we
briefly review derivation of the cosmological field equations when
the entropy associated with the apparent horizon of the universe
is in the form of Barrow entropy.

We take the line elements of the metric, for a homogeneous and
isotropic universe, as
\begin{equation}
ds^2=-dt^2+a{(t)}^2\left[\frac{dr^2}{1-kr^2}+r^2(d\theta^2+\sin^2
(\theta)d\varphi^2)\right],
\end{equation}
where $a(t)$ is the scale factor $k$ denotes the curvature of the
3 dimensional space. We assume the apparent horizon is the
boundary of the universe, with radius \cite{Sheykhi5}
\begin{align}
 \tilde{r}_{A}=\frac{1}{\sqrt{H^{2}+k / a^{2}}}.
\end{align}
This choice is consistent with first and second law of
thermodynamics. The work on the system is proportional to the
change in the volume ($dV$) and the first law on the apparent
horizon can be written as
\begin{equation}\label{FLaw}
dE=TdS+WdV,
\end{equation}
where the associated temperature to the apparent horizon is
\cite{Cai2,Sheyem}
\begin{equation}\label{temp}
T=-\frac{1}{2\pi\tilde{r}_A}\left(1-\frac{\dot{\tilde{r}}_A}{2H\tilde{r}_A}\right).
\end{equation}
while the work density is $W=\frac{1}{2}(\rho-p)$  \cite{hayward}.
Here $\rho$ and $p$ denote energy density and the pressure of the
cosmic fluids, respectively. Differentiating the total matter and
energy ($E=\rho V$) inside a $3$-dimensional sphere of radius
$\tilde{r}_{A}$, after using the continuity equation,
$\dot{\rho}+3H(\rho+p)=0$, we find
\begin{equation}\label{der}
dE=4\pi\tilde
 {r}_{A}^{2}\rho d\tilde {r}_{A}-4\pi H \tilde{r}_{A}^{3}(\rho+p) dt.
\end{equation}
Taking differential of the Barrow entropy expression
(\ref{bentropy}), we find
\begin{eqnarray} \label{dS}
dS &=&(2+\delta)\left(\frac{4\pi}{A_{0}}\right)^{1+\delta/2}
 {\tilde
{r}_{A}}^{1+\delta} \dot{\tilde {r}}_{A} dt.
\end{eqnarray}
Combining all equations with the first law of thermodynamics
(\ref{FLaw}), after some calculations, we reach \cite{Sheykhi5}
\begin{equation} \label{Fried2}
-\frac{2+\delta}{2\pi A_0
}\left(\frac{4\pi}{A_0}\right)^{\delta/2} \frac{d\tilde
{r}_{A}}{\tilde {r}_{A}^{3-\delta}}=
 \frac{1}{3}d\rho.
\end{equation}
Integration yields
\begin{equation} \label{Frie3}
\frac{2+\delta}{2-\delta}\left(\frac{4\pi}{A_0}\right)^{\delta/2}
\frac{1}{2\pi A_0} \frac{1}{\tilde
{r}_{A}^{2-\delta}}=\frac{\rho}{3},
\end{equation}
Substituting $\tilde {r}_{A}$, we find
\begin{equation} \label{Fried4}
\left(H^2+\frac{k}{a^2}\right)^{1-\delta/2} = \frac{8\pi G_{\rm
eff}}{3} \rho,
\end{equation}
where we have defined the effective Newtonian gravitational
constant as
\begin{equation}\label{Geff}
G_{\rm eff}\equiv \frac{A_0}{4} \left(
\frac{2-\delta}{2+\delta}\right)\left(\frac{A_0}{4\pi
}\right)^{\delta/2}.
\end{equation}
Equation (\ref{Fried4}) is the modified Friedmann equation based
on the Barrow entropy. Let us note that, the LHS of the Friedmann
equation get modified due to the modification in the entropy. This
is a reasonable result, since entropy is a geometry quantity and
thus any modification of it, should modify the geometry (gravity)
part of the field equations. This is one of the main difference
between our derivation and that of Ref. \cite{Saridakis1}, where
the authors modify the total energy density in the Friedmann
equations by considering the contribution of the Barrow entropy as
an effective DE in the RHS of field equations. In the limiting
case where $\delta=0$, the area law of entropy is recovered and we
have $A_{0}\rightarrow4G$. In this case, $G_{\rm eff}\rightarrow
G$, and Eq. (\ref{Fried4}) reduces to the standard Friedmann
equation in Einstein gravity.
\section{The Model} \label{Model}
The modified Friedmann equation in a flat universe is
\begin{equation}\label{fk0}
   H^{2-\delta}=\dfrac{1}{3M_{\mathrm{eff}}^{2}}(\rho_{m}+\rho_{de}),
\end{equation}
where $\rho_m$ is the energy density of a pressureless matter and
$\rho_{de}$ is energy density of the DE component. Using Eqs.
\eqref{lgo} and \eqref{rh}, the energy density of BHDE  with GO
cutoff can be written as
\begin{equation}\label{rde}
\rho_{de}=3M_{\mathrm{eff}}^2(\alpha
H^2+\beta\dot{H})^{1-\delta/2},
\end{equation}
where for latter convenience we have chosen
$C=3M_{\mathrm{eff}}^2$, Note that we can set $c^2=1$ without lose
of generality. When $\delta=0$ the energy density of HDE with GO
cutoff in standard cosmology is recovered. Given the modified
Friedmann equation \eqref{fk0} and the modified BHDE density
\eqref{rde} at hand, we will examine the evolution of the universe
from early decerebration to the late time acceleration.
\subsection{Dark energy dominated universe}
As a special case, let us consider the late time universe where
the DE is dominated. Thus, we can neglect the contribution from
matter and radiation. In this case the first Friedmann Eq.
\eqref{fk0} takes the following form
\begin{equation}
    H^{2-\delta}=\dfrac{\rho_{de}}{3M_{\mathrm{eff}}^{2}}.
\end{equation}
Combining with Eq. \eqref{rde}, the first modified Friedmann
equation through Barrow entropy reads
\begin{align}
 H^2=\alpha H^2+\beta\dot{H},
\end{align}
which admits the following solution for the Hubble parameter
\begin{align}
H(t)=\dfrac{\beta}{\alpha-1}\dfrac{1}{t}.
\end{align}
Therefore, the resulting scale factor is  $a(t)\propto
t^{\beta/(\alpha-1)}$. Using the continuity equation for BHDE as
$\dot{\rho}_{de}+3H\rho_{de}(1+w_{de})=0$ and noting that $w_{de}
= p_{de}/\rho_{de}$, we find the EoS parameter as
\begin{align}
    w_{de}=-1+\dfrac{2-\delta}{3}\left(\dfrac{\alpha-1}{\beta}\right).
\end{align}
Setting $\delta=0$, the result of HDE with GO cutoff in standard
cosmology is recovered \cite{Oliveros}. The deceleration parameter
is also written in the form
\begin{align}
 q=-1+\dfrac{\alpha-1}{\beta}.
\end{align}
It is seen that the Hubble parameter, $w_{de}$ and the
deceleration parameter $q$ only depend on $\alpha$, $\beta$ and
$\delta$. One can easily check that according to estimated range
of $\alpha$ and $\beta$ \cite{Manoharan:2024thb}, the fate of the
universe will be a big rip singularity in this case.
\subsection{Non-Interacting case}
\begin{figure*}[!ht]
 \centering
\includegraphics[trim={0cm 0cm 0cm 0.27cm},clip,scale=0.65]{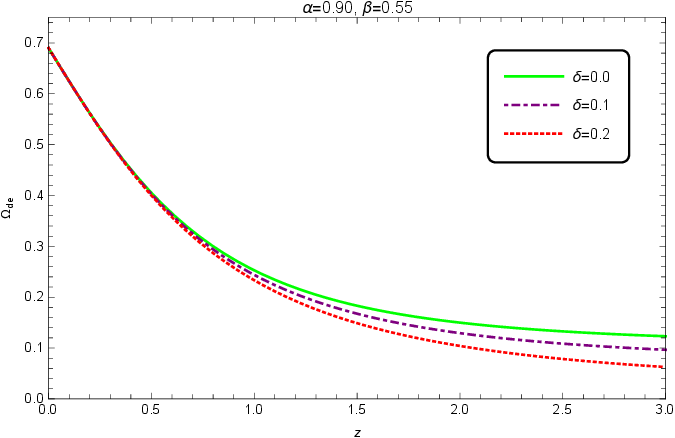}
\includegraphics[trim={0cm 0cm 0cm 0.cm},clip,scale=0.65]{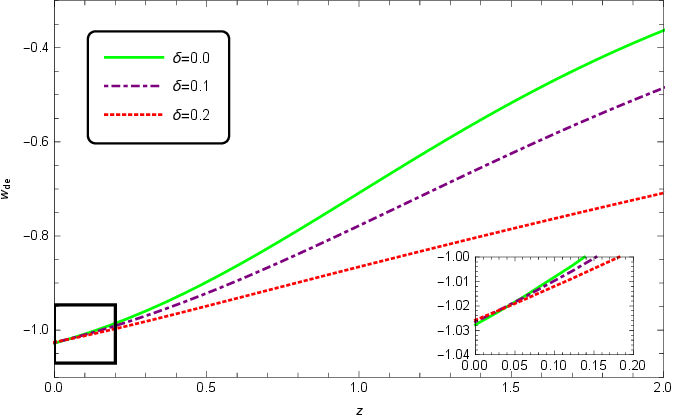}
\caption{\scriptsize Evolution of the fractional energy density
$\Omega_{de}$(the left panel) and the DE-EoS parameter,
$w_{de}$(the right panel) as a function of redshift, $z$, for
non-interacting HDE in a flat Barrow cosmology. Here, we set
$\Omega_{de,0}$= 0.69, $\alpha=0.93$ and
$\beta=0.55$.}\label{fig1}
\end{figure*}

In previous subsection we explored BHDE model for DE dominated
epoch. Here we would like to consider BHDE model in an epoch which
the cosmic fluids includes DM and DE components simultaneously. At
first we assume that the cosmic component are separately
conserved. The conservation equations when matter and DE evolve
separately read
\begin{equation}\begin{aligned}\label{pdd}
        \dot{\rho}_{de}+3H\rho_{de}(1+w_{de})=0,
\end{aligned}\end{equation}
\begin{equation}\begin{aligned}\label{pmm}
        \dot{\rho}_{m}+3H\rho_{m}=0.
\end{aligned}\end{equation}
The fractional density parameters are defined as usual
\begin{align}\label{omg}
    \Omega_{m}
    =\dfrac{\rho_{m}}{\rho_{cr}}=\dfrac{\rho_{m}}{3M_{\mathrm{eff}}^2H^{2-\delta}},\;\;\;\;\;\Omega_{de}
    =\dfrac{\rho_{de}}{\rho_{cr}}=\dfrac{\rho_{de}}{3M_{\mathrm{eff}}^2H^{2-\delta}},
\end{align}
where $\rho_{cr}=3M_{\mathrm{eff}}^2H^{2-\delta}$ is the effective
critical energy density. According to definition of \eqref{omg},
the first modified Friedman equation \eqref{fk0} deforms to
\begin{align}
    \Omega_{m}+\Omega_{de}=1.
\end{align}
Substituting $\dot{\rho}_{m}$ and $\dot{\rho}_{de}$ from the
conservation equations \eqref{pdd} and \eqref{pmm} into time
derivative of the modified Friedmann equation \eqref{fk0}, we
arrive at
\begin{align}\label{dhbh}
\dfrac{\dot{H}}{H^2}=\dfrac{-3}{( 2-\delta)}\left[1+ w_{de}
\Omega_{de}\right].
\end{align}
Dividing Eq.\eqref{fk0} by $\rho_{de}$ and using relation
\eqref{rde}, we find
\begin{equation} \label{dhbh2}
    \left( \alpha+\beta\dfrac{\dot{H}}{H^2}\right) ^{(\delta-2)/2}=\left(1+\dfrac{\rho_{m}}{\rho_{de}} \right).
\end{equation}
Substituting Eq.\eqref{dhbh} into Eq. \eqref{dhbh2}, we find the
EoS parameter $w_{de}$ as
\begin{align}\label{wd}
w_{de}=\dfrac{1}{\Omega_{de}}\left[ \dfrac{( \delta-2)}{3\beta}(\Omega_{de}^{2/(2-\delta)}-\alpha)-1\right].
\end{align}
In order to explore the behaviour of the DE model during the
history of the universe, one should consider the evolution of DE
density parameter $\Omega_{de}$. Taking derivative of
$\Omega_{de}$ with respect to redshift parameter $z$ yields
\begin{align}\label{kkk}
    \dfrac{d\Omega_{de}}{dz}=\dfrac{d\Omega_{de}}{dt}\dfrac{dt}{dz}=\dfrac{d}{dt} \left(\dfrac{\rho_{de}}{3M^2_p H^{2-\delta}}\right)\left(-\dfrac{1}{H(1+z)}
    \right).
\end{align}
After a little algebra and using Eq.\eqref{pdd} one gets
\begin{align}\label{u}
    \dfrac{d\Omega_{de}}{dz}=\dfrac{3\Omega_{de}w_{de}(1-\Omega_{de})}{(1+z)}.
\end{align}
Using Eq.\eqref{wd}, we arrive at
\begin{align}\label{omgd}
\dfrac{d\Omega_{de}}{dz}=\dfrac{3(1-\Omega_{de})}{(1+z)}\left[
\dfrac{(\delta-2)}{3\beta}(\Omega_{de}^{2/(2-\delta)}-\alpha)-1\right].
\end{align}
We plot the evolution of $\Omega_{de}$ and $w_{de}$ with respect
to redshift for BHDE with GO IR cutoff in  Fig.\ref{fig1} for
different values of Barrow exponent $\delta$. From this figure, we
observe that for $(z>0.6)$, the value of $\Omega_{de}$ decreases
with the increase of $\delta$ while for $0<z<0.6$ there is no
dependency to $\delta$. Also the difference between $\Omega_{de}$
in standard cosmology ($\delta=0$) and modified Barrow cosmology
has been larger in the past. It should be noted that the
parameters $\alpha$ and $\beta$ are in the range of $[0.7,1.0]$
and $[0.3,0.9]$ \cite{Manoharan:2024thb}, respectively, and we
find that the best agreement for our model achieves for
$\alpha=0.93$ and $\beta=0.55$. Thus in the rest of this paper we
will use these values of $\alpha$ and $\beta$.

Let us probe in more detail the EoS parameter in Fig.\ref{fig1}. A
overall harvest is that by increasing the Barrow exponent $\delta$, the EoS parameter  $w_{de}$ decreases. According to this figure, the EoS parameter at
present time crosses the phantom divide $(w_{de}<-1)$. It is worth
noting that for $0\leq z<0.1$, the $w_{de}$ lies in the phantom
regime and for larger $\delta$, the phantom
divide is crossed earlier. One should note that this behavior
is cannot be seen in the non-interacting BHDE with future event
horizon as IR cutoff \cite{SheBHDE}.

\begin{figure}[!ht]
\centering
\includegraphics[trim={0cm 0cm 0cm 0.27cm},clip,scale=0.65]{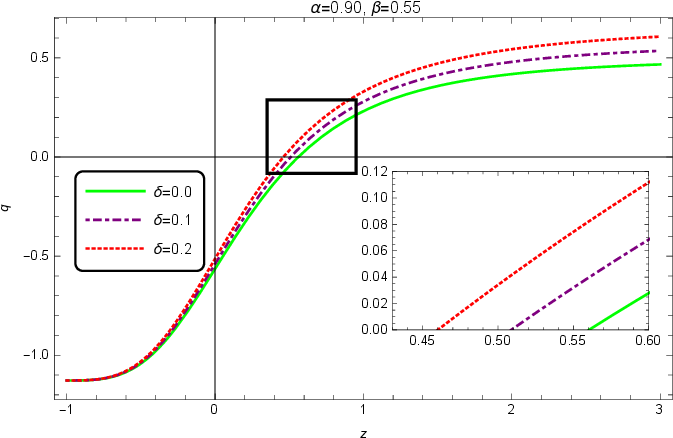}
\caption{\scriptsize Evolution of the deceleration parameter  $q$
against $z$ in a non-interacting BHDE. Here, we set $\Omega_{de,0}$= 0.69, $\alpha=0.93$ and
$\beta=0.55$.}\label{fig3}
\end{figure}

The deceleration parameter is obtained as
\begin{align}\label{q1}
q=-1-\dfrac{\dot{H}}{H^2}=-1+\dfrac{3}{( 2-\delta)}\left[1+ w_{de}
\Omega_{de}\right]=-1-
\dfrac{\Omega_{de}^{2/(2-\delta)}-\alpha}{\beta}.
\end{align}
The evolution of the deceleration parameter in terms of the
redshift parameter $z$ is presented in Fig.\ref{fig3}. According
to this figure, we find that by increasing the Barrow exponent,
$\delta$, the transition from a decelerated universe $(q>0)$ to an
accelerated universe $(q<0)$ happens in the lower redshifts. For
example for $\delta=0.1$, the phase transition occurs at $z=0.51$.
Thus increasing $\delta$ parameter leads to a delay in the cosmic
phase transition during the history of the universe.
\subsection{Interacting case}
Next we analyze a FRW universe filled by BHDE and DM, while
swapping energy among them. Since the microscopic nature of both
DE and DM is still unknown, there exist enough room to leave a
chance of interaction between DE and DM \cite{wetter}. Beside it
is shown that in the presence of interaction between two dark
components the coincidence problem could be alleviated
\cite{interact1,oli}. In \cite{Pereira}, the interaction between
DM and DE has been extensively investigated from a thermodynamic
point of view. Thus there exist enough motivation and interest to
explore an interacting version of dark components. The
semi-conservation equations including the interaction between DM
and DE read
\begin{equation}\begin{aligned}\label{pd}
        \dot{\rho}_{de}+3H\rho_{de}(1+w_{de})=-Q,
\end{aligned}\end{equation}
\begin{equation}\begin{aligned}\label{pm}
        \dot{\rho}_{m}+3H\rho_{m}=Q,
\end{aligned}\end{equation}
where $Q=\Gamma\rho_{de}=3b^2H(1+r)\rho_{de}$ is the interaction
term, with $b^2$ is  a coupling constant and
$r=\rho_{m}/\rho_{de}$ \cite{Pavon}. The negative sign of $Q$
indicates a transfer of energy from DE to DM. The interaction
between dark components affects evolution of the model. Following
a same steps as the previous subsection one can find the resulting
equations for $w_{de}$, $q$ as well as the evolution equation of
$\Omega_{de}$. At first step Substituting $\dot{\rho}_{m}$ and $\dot{\rho}_{de}$ from the
conservation equation \eqref{pd} and \eqref{pm} into time derivative of modified Friedmann equation \eqref{fk0}, we find
\begin{align}\label{eh}
    \dfrac{\dot{H}}{H^2}=\dfrac{-3}{( 2-\delta)}\left[1+ w_{de}
    \Omega_{de}\right].
\end{align}
In the next step, dividing the modified Friedmann equation in a flat universe \eqref{fk0} by $\rho_{de}$ and according to Eq.\eqref{eh}, we arrived the EoS parameter $w_{de}$ as
\begin{align}
    w_{de}=\dfrac{1}{\Omega_{de}}\left[ \dfrac{( \delta-2)}{3\beta}(\Omega_{de}^{2/(2-\delta)}-\alpha)-1\right].
\end{align}
Then we obtain evolution equation of the HDE in Barrow cosmology, This equation versus redshift reads
\begin{align}
\dfrac{d\Omega_{de}}{dz}=\dfrac{3}{(1+z)}\left[(1-\Omega_{de})\left(
\dfrac{(
\delta-2)}{3\beta}(\Omega_{de}^{2/(2-\delta)}-\alpha)-1\right)+b^2
\right].
\end{align}
Also, the deceleration parameter can be written as
\begin{align}\label{q1}
    q=-1-\dfrac{\dot{H}}{H^2}=-1+\dfrac{3}{( 2-\delta)}\left[1+ w_{de}
    \Omega_{de}\right]=-1-
    \dfrac{\Omega_{de}^{2/(2-\delta)}-\alpha}{\beta}.
\end{align}
We found the EoS parameter $w_{de}$ and the deceleration parameter  $q$ in this case and equations \eqref{wd} and \eqref{q1} are reconstructed. Therefore, the existence of interaction appeared in the evolution of the fractional density parameter against redshift.

\begin{figure*}[!ht]
\centering
\includegraphics[trim={0cm 0cm 0cm 0.27cm},clip,scale=0.65]{{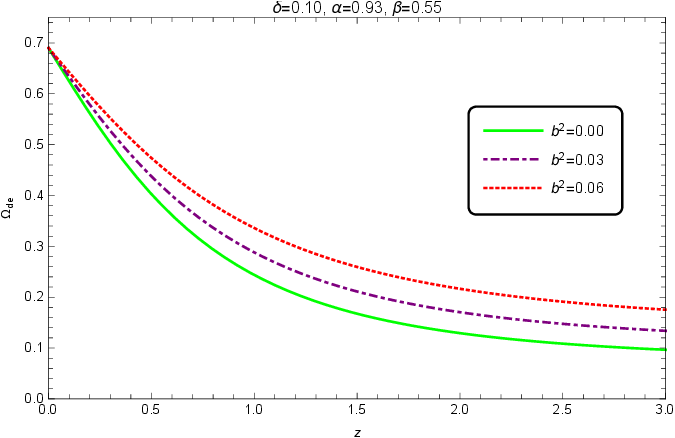}}
\hspace{5mm}
\includegraphics[trim={0cm 0cm 0cm 0.31cm},clip,scale=0.65]{{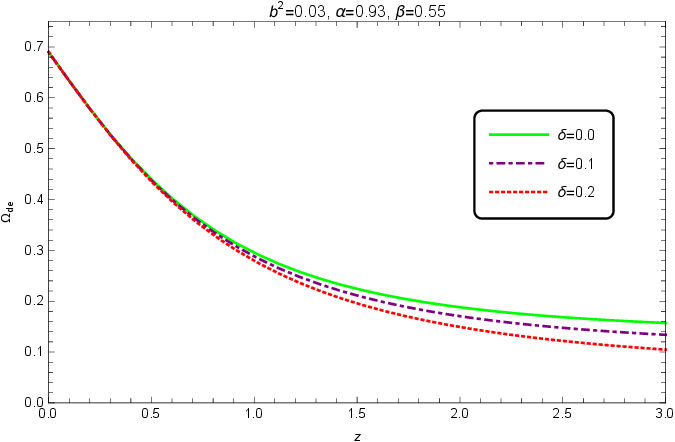}}
\caption{\scriptsize Evolution of $\Omega_{de}$ against $z$ for
HDE in a flat Barrow cosmology for different values of $b^2$(left)
and $\delta$ (right). Here, we have set $\Omega_{de,0}$=0.69, $\delta=0.1$(in
left panel),$b^2=0.03$ (in the right panel), $\alpha=0.93$ and $\beta=0.55$.}
 \label{f1}
\end{figure*}

\begin{figure*}[!ht]
\centering
 \includegraphics[trim={0cm 0cm 0cm 0cm},clip,scale=0.65]{{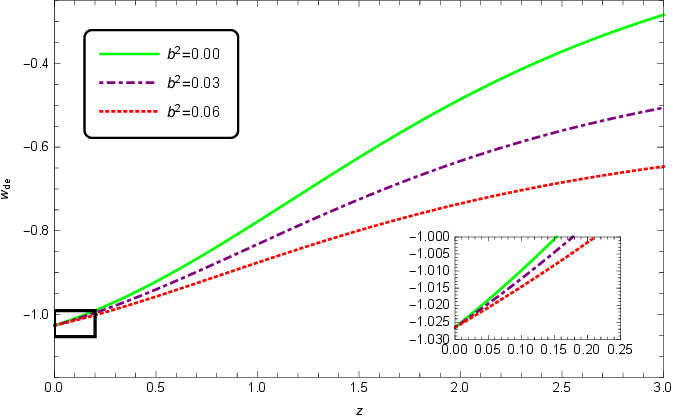}}
  \hspace{5mm}
  \includegraphics[trim={0cm 0cm 0cm 0cm},clip,scale=0.65]{{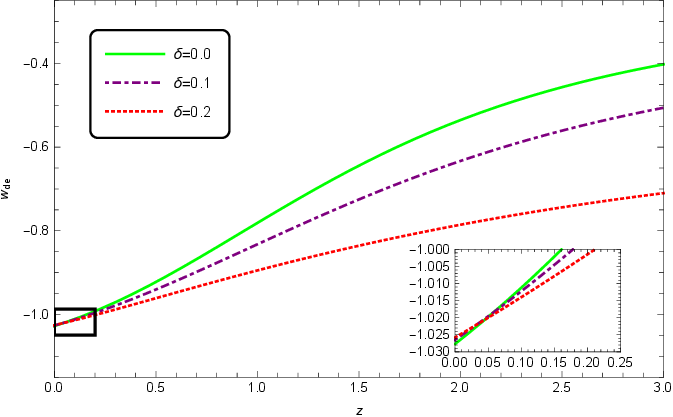}}
\caption{\scriptsize Evolution of $w_{de}$ against redshift for
flat interacting BHDE in Barrow cosmology for different choices of $b^2$ (left) and $\delta$ (right). In above figures, we set
$\Omega_{de,0}$= 0.69, $\delta=0.1$(in
left panel),$b^2=0.03$ (in the right panel), $\alpha=0.93$ and
$\beta=0.55$.} \label{f2}
\end{figure*}

\begin{figure*}[!ht]
\centering
 \includegraphics[trim={0cm 0cm 0cm 0.27cm},clip,scale=0.65]{{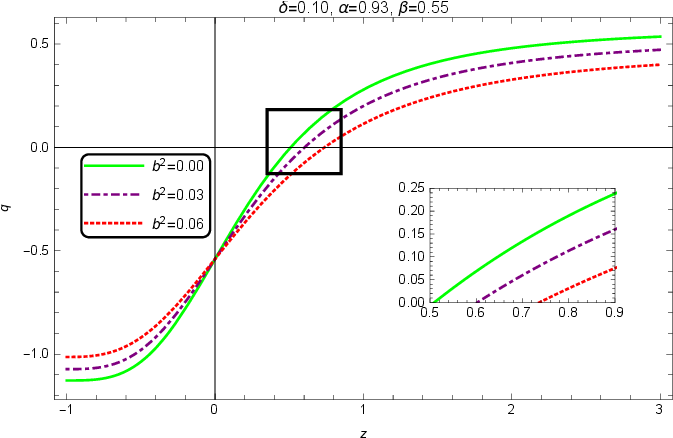}}
 \hspace{5mm}
 \includegraphics[trim={0cm 0cm 0cm 0.31cm},clip,scale=0.65]{{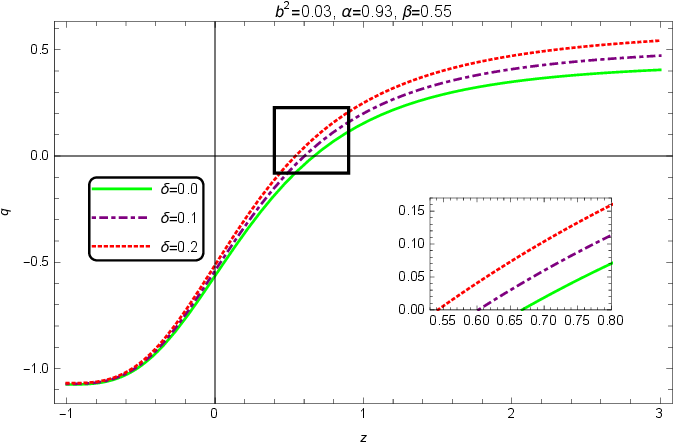}}
\caption{\scriptsize Evolution of the deceleration parameter($q$)
against redshift in flat interacting HDE in Barrow cosmology for
different choices of $b^2$(left) and $\delta$(right). In above
figures, we set $\Omega_{de,0}$= 0.69, $\delta=0.1$(in
left panel),$b^2=0.03$ (in the right panel), $\alpha=0.93$ and
$\beta=0.55$.} \label{f3}
\end{figure*}

The evolution of $\Omega_{de}$ versus redshift for different
values of $b^2$ and $\delta$ is plotted in Fig.\ref{f1}.
Notice that at each $z$, $\Omega_{de}$ decreases as $b^2$
decreases. From right panel of Fig.\ref{f1} one can find that
Barrow exponent $\delta$ can affects the universe at earlier
epochs while its impact at present epoch it has no significant
effect.
Next we turn to $w_{de}$ as well as $q$ for a more deep
description of the model. At first, one can easily see that
$w_{de}$ and $q$ in the interacting HDE of interest reduce to the
corresponding relations in the non-interacting case by setting
$b^2=0$. A close look to left panel of Fig.\ref{f2} reveals that
with increasing the coupling constant $b^2$, the EoS parameter
decreases as well. For all choices of $b^2$, $w_{de}$ lies in the
phantom regime at the present epoch.
In the left plot of Fig.\ref{f3}, the evolution of the
deceleration parameter $q$ as a function of redshift is shown for
different values of coupling constant $b^2$. For $b^2=0.03$, the
phase transition occurs at $z=0.6$ while for $b^2=0.06$, the phase
transition occurs at $z=0.74$. Despite the effects of $\delta$,
increasing $b^2$ causes an earlier phase transitions.

\section{HDE in nonflat universe} \label{Nonflat}
\begin{figure}[!ht]
 \centering
\includegraphics[trim={0cm 0cm 0cm 0cm},clip,scale=0.7]{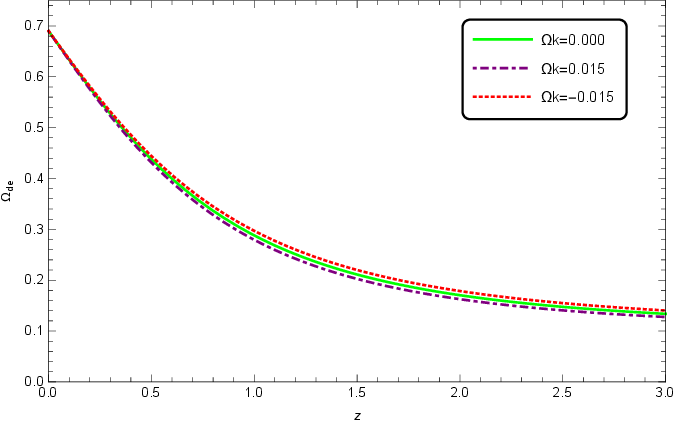}
\caption {\scriptsize Evolution of  $\Omega_{de}$ against $z$, in
non-flat and flat cases. Here, we also set
$\Omega_{de,0}$= 0.69, $b^2=0.03$, $\delta=0.1$, $\alpha=0.93$ and
$\beta=0.55$.}\label{fig10}
\end{figure}

\begin{figure}[!ht]
\centering
 \includegraphics[trim={0cm 0cm 0cm 0cm},clip,scale=0.7]{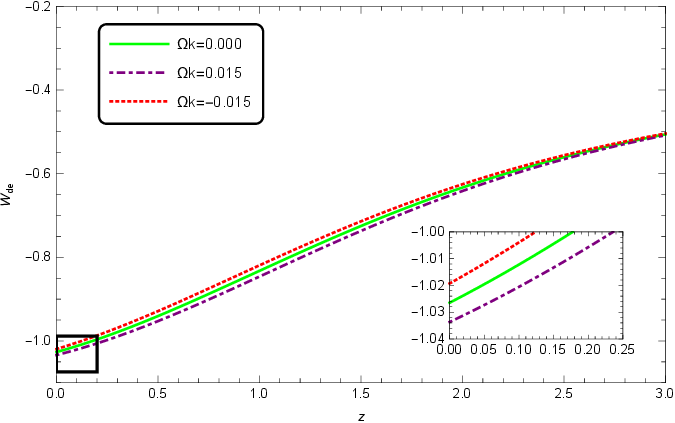}
 \caption
 {\scriptsize Evolution of  $w_{de}$ with respect to $z$, in non-flat and flat cases. Here, we also set $\Omega_{de,0}$= 0.69, $b^2=0.03$, $\delta=0.1$, $\alpha=0.93$ and $\beta=0.55$.}\label{fig11}
\end{figure}

\begin{figure}[!ht]
    \centering
    \includegraphics[trim={0cm 0cm 0cm 0cm},clip,scale=0.7]{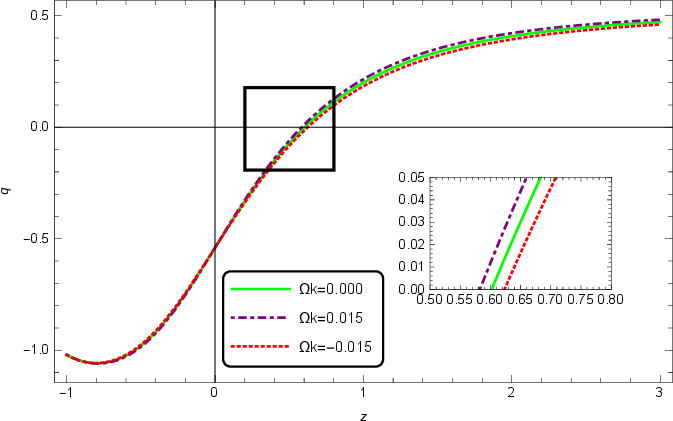}
    \caption{\scriptsize  Evolution of the deceleration parameter, $q$, as against $z$ in non-flat and flat, interacting HDE in Barrow cosmology. Here, we set $\Omega_{de,0}$= 0.69, $b^2=0.03$, $\delta=0.1$, $\alpha=0.93$ and $\beta=0.55$.}\label{fig12}
\end{figure}

There exist observational evidences reveal that our observed
Universe is not exactly flat and there exist signs of a nonflat
geometry \cite{Bennett}-\cite{Spergel2}. Thus, it is well
motivated to consider dynamics of a nonflat universe. In order to
study the dynamics of a nonflat universe filled with BHDE and DM,
we introduce the corresponding curvature density parameter,
$\Omega_{k}=\dfrac{k}{a^2 H^2}$. The first modified Friedmann
equation through Barrow entropy in a nonflat universe is
\begin{equation}\label{i}
3M_{\mathrm{eff}}^{2}\left(H^{2}+\dfrac{k}{a^{2}}\right)
^{1-\delta/2}=\rho_{m}+\rho_{de}.
\end{equation}
Taking the time derivative of the above equation and using
$\dot{\rho}_{m}$ and $\dot{\rho}_{de}$ from the continuity
equations Eqs.\eqref{pm} and \eqref{pd}, we find
\begin{equation}\label{x}
\dfrac{\dot{H}}{H^{2}} =\dfrac{-3}{2-\delta}\left(
1+\dfrac{w_{de}}{1+r}\right) -\dfrac{3}{2-\delta}\Omega_{k}\left(
1+\dfrac{w_{de}}{1+r}\right) +\Omega_{k}.
\end{equation}
As mentioned, $r=\rho_{m}/\rho_{de}$ is the ratio of DM density to
DE density. Dividing Eq.\eqref{i} by $\rho_{de}$ and substituting
Eq.\eqref{rde}, we arrive at
\begin{equation}
    1+r=\dfrac{\left(1+\dfrac{k}{a^{2}H^{2}}\right)^{1-\delta/2}}{\left(\alpha +\beta \dfrac{\dot{H}}{H^{2}}\right)^{1-\delta/2}}.
\end{equation}
Substituting Eq.\eqref{x} and after a little algebra and mixing the related relations, we find
\begin{align}
w_{de}=&\dfrac{(1+\Omega_{k})^{-\delta/2}}{\Omega_{de}}\left\lbrace
-\dfrac{2-\delta}{3\beta}\left[
\Omega_{de}^{2/(2-\delta)}-\left(\alpha-\dfrac{2\beta}{2-\delta}\right)
+\dfrac{\beta}{2-\delta}\left((1+\delta)\Omega_{k}+1
\right)\right] \right\rbrace.
\end{align}
One can easily check that this equation reduce to the
corresponding one in the flat background for $k=0$. Evolution of
the fractional energy densities versus redshift according to
$\dfrac{d\Omega_{i}}{dz}=\dfrac{d\Omega_{i}}{dt}\dfrac{dt}{dz}$
and Eq.\eqref{pd} are obtained as
\begin{align}
 \nonumber
 \dfrac{d\Omega_{de}}{dz}
=&\dfrac{\Omega_{de}}{(1+z)}\biggl[3(1+w_{de})+3b^{2}\dfrac{(1+\Omega_{k})^{1-\delta/2}}{\Omega_{de}}+(2-\delta)\biggl[-\dfrac{3}{2-\delta}\left(1+\dfrac{w_{de}\Omega_{de}}{(1+\Omega_{k})^{1-\delta/2}}\right)\\&
-\dfrac{3}{2-\delta}\Omega_{k}\left(1+\dfrac{w_{de}\Omega_{de}}{(1+\Omega_{k})^{1-\delta/2}}\right)+\Omega_{k}\biggr]
\biggr].
\end{align}
\begin{align}
    \dfrac{d\Omega_{k}}{dz}=
    -2\dfrac{\Omega_{k}}{(1+z)}\left\lbrace -1+\dfrac{3}{2-\delta}\left(1+\dfrac{w_{de}\Omega_{de}}{(1+\Omega_{k})^{1-\delta/2}}\right)+\dfrac{3}{2-\delta}\Omega_{k}\left(1+\dfrac{w_{de}\Omega_{de}}{(1+\Omega_{k})^{1-\delta/2}} \right)-\Omega_{k}\right\rbrace.
\end{align}
Since the relations are messy analytic description of the model
look hard and thus we disclose different features of the model
through numerical analysis. On this way we plotted evolution of
$\Omega_{de}$ against redshift in open, flat and close geometry in
Fig.\ref{fig10}.
Evolution of the EoS parameter for interacting BHDE in Barrow
cosmology in non-flat and flat universes are shown in
Fig.\ref{fig11}. For all choices of the free parameters $w_{de}$
crosses the phantom regime at present epoch in agreement with
observations.
According to obtained term for $\dot{H}/H^2$ in non-flat universe
Eq.\eqref{x}, the deceleration parameter is written as
\begin{align}
    q=-1+\dfrac{3}{2-\delta}\left( 1+\dfrac{w_{de}}{1+r}\right) +\dfrac{3}{2-\delta}\Omega_{k}\left( 1+\dfrac{w_{de}}{1+r}\right) -\Omega_{k}.
\end{align}
The evolution of the deceleration parameter versus redshift in
presence of interaction among dark sector components in non-flat
and flat universe are presented in Fig.\ref{fig12}. From this
figure, we see that for open universe, the transition from
deceleration $(q>0)$ to acceleration $(q<0)$ happened at lower
redshift $(z=0.58)$ and for closed universe, the transition from
deceleration $(q>0)$ to acceleration $(q<0)$ happened at higher
redshift $(z=0.62)$ also for flat universe, the transition
happened at $z=0.6$. These indicates that all cases are consistent
with observations, $z_t\in[0.5,0.7]$ \cite{komatsu} depending on
suitable choices of the free parameters.

\section{Stability and Statefinder of the model}\label{Stab}
\subsection{Stability}
\begin{figure}[!ht]
    \centering
    \includegraphics[trim={0cm 0cm 0cm 0cm},clip,scale=0.625]{{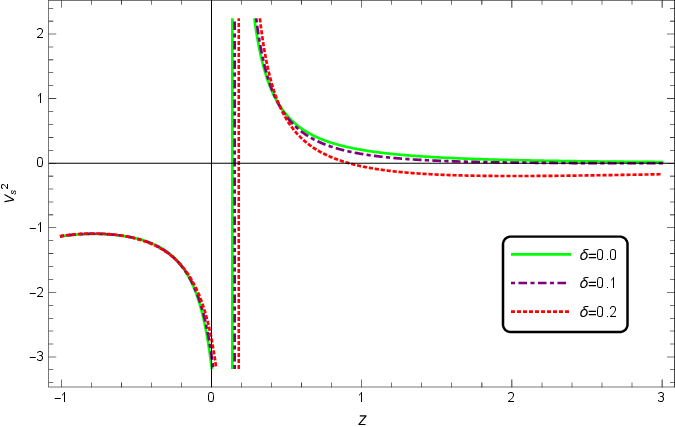}}(a)
    \includegraphics[trim={0cm 0cm 0cm 0cm},clip,scale=0.625]{{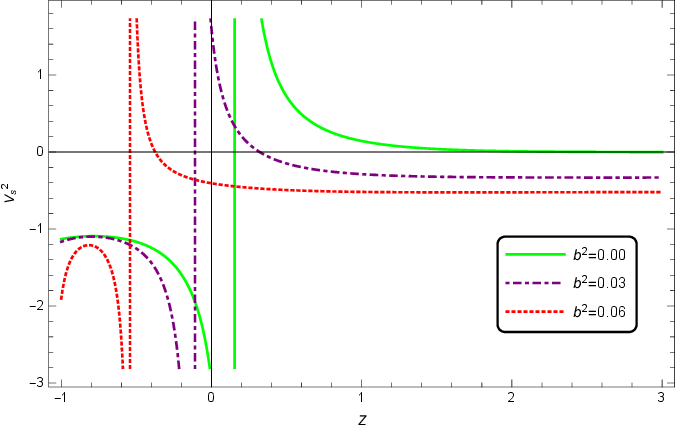}}(b)
    \includegraphics[trim={0cm 0cm 0cm 0cm},clip,scale=0.625]{{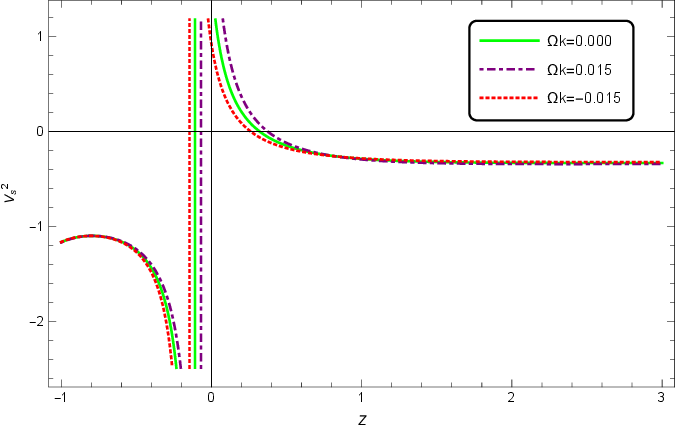}}(c)
    \caption{\scriptsize (a) The squared sound speed $v^2_s$ as a function of the redshift z in flat and non-interacting HDE, in Barrow cosmology  for different values of $\delta$. Here, we have set  $\Omega_{de,0}$= 0.69,  $\alpha=0.93$ and $\beta=0.55$. (b) \scriptsize The square sound speed for different choices of $b^2$. Here, we set $\delta=0.1$ and the other parameters are the same as panel (a). (c) The square sound speed versus $z$ in  flat and non-flat geometries. In this case we set $b^2=0.03$, $\delta=0.1$ while the other constants are the same as other panels.}
\label{f4}
\end{figure}

Squared sound speed($v^{2}_s$) analysis, is an interesting method
which reveals sounds of gravitational instability through a
semi-Newtonian approach. In an expanding background, deep inside
the horizon, one can find evolution of the cosmic perturbations as
\begin{equation}\label{delta}
\delta\propto e^{\pm i\omega t},
\end{equation}
where $\omega\propto v_s$. The negative and  positive signs of
$v_s^2$ could be a sound of instability and stability
respectively. To proceed the analysis, we start with definition of
the adiabatic squared sound speed which reads

\begin{equation}\label{vs2}
    v_{s}^{2}=\frac{dp_{de}}{d\rho_{de}},
\end{equation}
where $p_{de}$ and $\rho_{de}$ are the pressure and the density of the DE component, respectively.
According to $p_{de}=\rho_{de}w_{de}$, Eq.\eqref{vs2} can be written as
\begin{align}\label{vs}
    v_{s}^{2}=\dfrac{\dot{\rho_{de}w_{de}}}{\dot{\rho}_{de}}
    =w_{de}+\dfrac{\rho_{de}}{\dot{\rho}_{de}}\dfrac{d{w}_{de}}{d\Omega_{de}}\dfrac{d\Omega_{de}}{dt}.
\end{align}
Substituting ${\rho_{de}}/{\dot{\rho}_{de}}$ from the conservation
equation \eqref{pd}, one may get
\begin{align}
    \nonumber  v_{s}^{2}=&w_{de}+\dfrac{d{w}_{de}}{d\Omega_{de}}\left\lbrace  \dfrac{\Omega_{de}}{-3H(1+w_{de})-3Hb^2\dfrac{(1+\Omega_{k})^{1-\delta/2}}{\Omega_{de}}}\right\rbrace  \biggl[3(1+w_{de})+3b^{2}\dfrac{(1+\Omega_{k})^{1-\delta/2}}{\Omega_{de}}\\
    &+(2-\delta)\biggl[-\dfrac{3}{2-\delta}\left(1+\dfrac{w_{de}\Omega_{de}}{(1+\Omega_{k})^{1-\delta/2}}\right) -\dfrac{3}{2-\delta}\Omega_{k}\left(1+\dfrac{w_{de}\Omega_{de}}{(1+\Omega_{k})^{1-\delta/2}}\right)+\Omega_{k}\biggr] \biggr].
\end{align}
Effects of $\delta$ and the coupling constant $b^2$ are shown in Fig.\ref{f4}. It can be seen that increasing of $\delta$, the
ability in the model decreases. It is interesting that $v_s^2$
takes positive values in the past and tends to positive values in
the future, Fig.\ref{f4}(a). In this model, evidences of
instability is seen in the presence of an interaction between DM
and DE in the past Fig.\ref{f4}(b). Also, the stability analysis
results of the flat, open and closed universe are almost similar.
$v_s^2$ is positive in the near past and near future
Fig.\ref{f4}(c).
\subsection{Statefinder}
\begin{figure}[!ht]
    \centering
    \includegraphics[trim={0cm 0cm 0cm 0cm},clip,scale=0.625]{{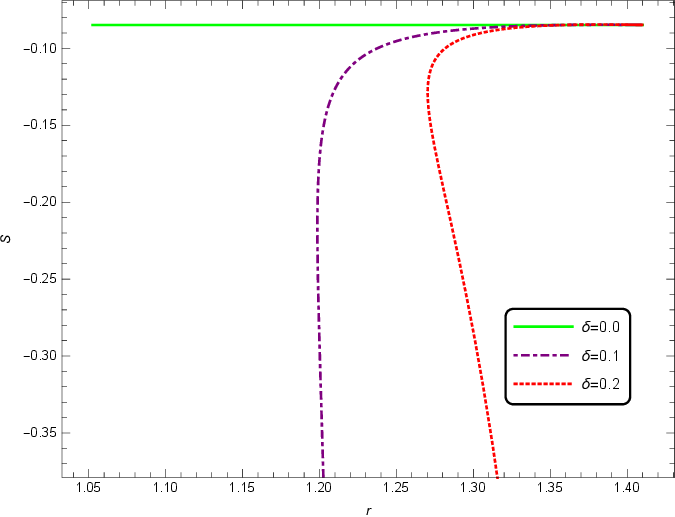}}(a)
    \includegraphics[trim={0cm 0cm 0cm 0cm},clip,scale=0.515]{{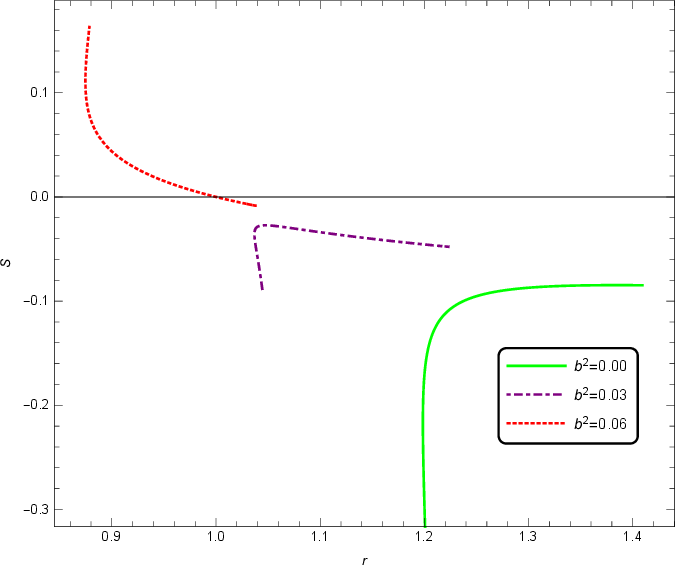}}(b)
    \includegraphics[trim={0cm 0cm 0cm 0cm},clip,scale=0.55]{{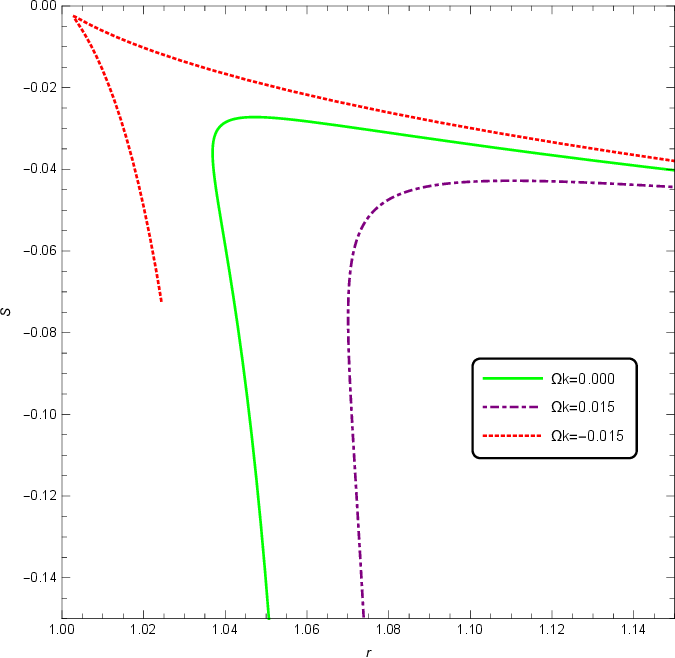}}(c)
    \caption{\scriptsize(a) The evolution of $s$ against $r$ in flat and non-interacting HDE in Barrow cosmology for different value of $\delta$. Here, we set  $\Omega_{de,0}$= 0.69, $\alpha=0.93$ and $\beta=0.55$. (b) A same plot as the panel (a) but in presence of interaction and a same choice of the free parameters (c) The evolution of $s$ versus $r$ in flat and non-flat HDE in Barrow cosmology. The free parameters are set to a same values.}
    \label{f5}
  \end{figure}

\begin{figure}[!ht]
    \centering
    \includegraphics[trim={0cm 0cm 0cm 0cm},clip,scale=0.625]{{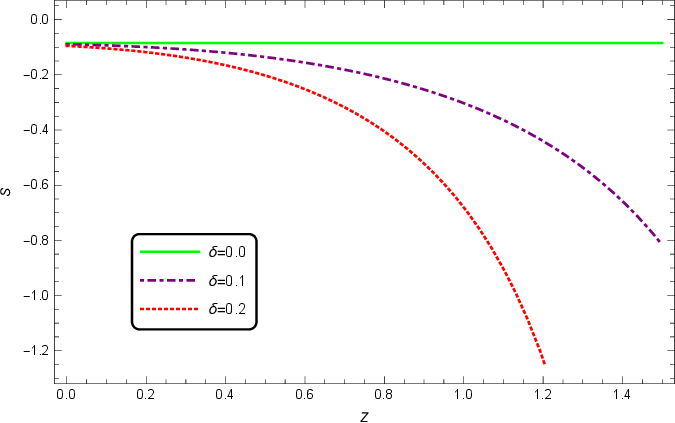}}(a)
    \includegraphics[trim={0cm 0cm 0cm 0cm},clip,scale=0.625]{{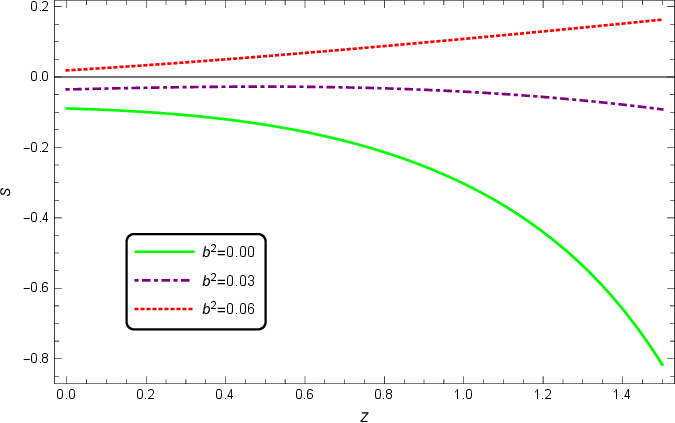}}(b)
    \includegraphics[trim={0cm 0cm 0cm 0cm},clip,scale=0.625]{{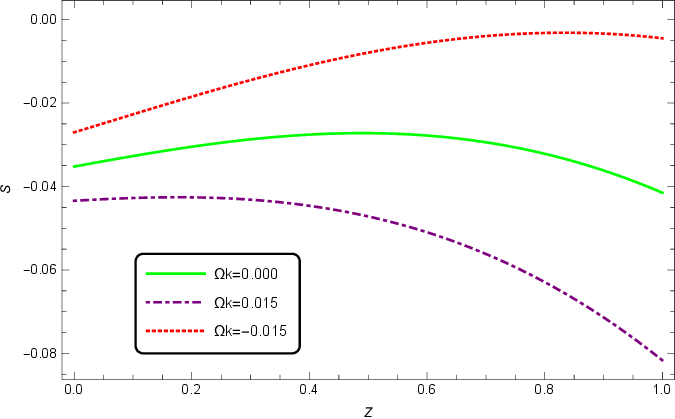}}(c)
    \caption{\scriptsize(a) The evolution of $s$ against $z$ in flat and non-interacting HDE in Barrow cosmology for different value of $\delta$. Here, we set $\Omega_{de,0}$= 0.69, $\alpha=0.93$ and $\beta=0.55$. (b) A same plot as the panel (a) but in presence of interaction and a same choice of the free parameters. (c) The evolution of $s$ versus $z$ in flat and non-flat HDE in Barrow cosmology. The free parameters are set to a same values.}
    \label{f6}
\end{figure}

Sahni et al. \cite{Sahni} introduced the statefinder pair
$\left\lbrace  r,s\right\rbrace$, which includes third time
derivative of the scale factor $a(t)$. Since DE models are very
diverse, it is useful and essential to distinguish between
different DE models. The statefinder pair $\left\lbrace
r,s\right\rbrace$ is an effective tool for finding differences
between DE models and the standard model($\Lambda$CDM). It should
be noted, for the standard model $\Lambda$CDM the statefinder
parameters are $\left\lbrace r,s\right\rbrace= \left\lbrace
1,0\right\rbrace$. The statefinder pair $\left\lbrace
r,s\right\rbrace$ is defined as \cite{Sahni},\cite{Alam}
\begin{align}\label{rr}
    r=\dfrac{\dddot{a}}{aH^3}=2q^2+q-\dfrac{\dot{q}}{H}.
\end{align}
\begin{align}\label{s}
    s=\dfrac{r-1}{3[q-(1/2)]}.
\end{align}
Based on the relationship between $r$ and the deceleration parameter $q$, the $r$ parameter could be obtained as
\begin{align}
    \nonumber
    r&=2q^2+q-\dfrac{1}{H}\dfrac{dq}{d\Omega_{de}}\Omega_{de}\biggl[3(1+w_{de})+3b^{2}\dfrac{(1+\Omega_{k})^{1-\delta/2}}{\Omega_{de}}+(2-\delta)\biggl[-\dfrac{3}{2-\delta}\left(1+\dfrac{w_{de}\Omega_{de}}{(1+\Omega_{k})^{1-\delta/2}}\right)\\
    &-\dfrac{3}{2-\delta}\Omega_{k}\left(1+\dfrac{w_{de}\Omega_{de}}{(1+\Omega_{k})^{1-\delta/2}}\right)+\Omega_{k}\biggr] \biggr].
\end{align}
We plotted the evolution of $s$ as a function of $z$ for different
value of $\delta$, $b^2$ and also for closed, open and flat
universe in Fig.\ref{f5} and Fig.\ref{f6}. As expected, increasing
$\delta$, the BHDE in Barrow cosmology model becomes more
different with respect to $\Lambda$CDM, which is evident in
Fig.\ref{f5}(a) and Fig.\ref{f6}(a). Also, by increasing $b^2$,
the statefinder parameters tend to the values $r\to 1$ and $s \to
0$ and in a special case for $b^2=0.06$, the statefinder
parameters catch the values $\left\lbrace  r,s\right\rbrace=
\left\lbrace 1,0\right\rbrace$ at the present time. This means
that the our interested HDE in Barrow cosmology model is close to
the $\Lambda CDM$ at the present time in the interacting case
according to Fig.\ref{f5}(b) and  Fig.\ref{f6}(b). In an open
universe the model evolves close to the values $r\to 1$ and
  $s \to 0$ in the past epochs which can be seen in Fig.\ref{f5}(c) and Fig.\ref{f5}(c).

\section{Conclusions} \label{Con}
Based on thermodynamics-gravity conjecture and holographic
principle, on the cosmological setups, any modification to the
entropy, not only change the energy density of the HDE but also
leads to the modified Friedmann equations. In most studies on HDE
through modified entropy expression, people consider a modified
expression for the HDE density, while they still use the usual
Friedmann equations in standard cosmology. This is not compatible
with thermodynamics-gravity conjecture, which implies that using
the first law of thermodynamics and modified entropy expression,
the cosmological field equations get modified as well.

In this work we have investigated BHDE model with GO cutoff in the
background of modified Barrow cosmology. We have examined several
cases, including DE dominated universe, a flat universe filled
with BHDE and DM which evolves separately, a flat universe filled
with interacting BHDE and DM. We find that, regardless of the
interaction term between two dark components, the EoS parameter,
$w_{de}$, decreases by increasing the Barrow exponent $\delta$
during the history of the universe, and crosses the phantom divide
at the present time. Also the Barrow exponent $\delta$ changes the
time of the universe phase transition. We observed that increasing
$\delta$ parameter leads to a delay in the cosmic phase transition
during the history of the universe. In other words, the phase
transition of the universe take place at lower redshift by
increasing $\delta$. Also for a fixed value of Barrow exponent
$\delta$, by increasing the interaction parameter $b^2$, the value
of EoS parameter decreases in all epochs.

We also generalized our studies to a nonflat universe. We found
that in flat and non-flat universe, the EoS parameter $w_{de}$
crosses the phantom regime and we saw that all cases are
compatible with observations, $z_t\in[0.5,0.7]$ \cite{komatsu}
depending on suitable choices of the free parameters.

We also explored the sound stability and statefinder of this model
and found out that by increasing of $\delta$, the stability of the
model decreases and this model becomes more different with respect
to $\Lambda$CDM.

\textbf{Acknowledgements:}{The work of A. Sheykhi is based upon
research funded by Iran National Science Foundation (INSF) under
project No. 4024978.}

\textbf{Competing interests:} The authors declare there are no
competing interests.

\textbf{Availability of data:}This manuscript does not report
data.


\begin{thebibliography}{99}
\bibitem{riess} A. G. Riess et al, \textit{Observational evidence from supernovae for an accelerating universe and a cosmological constant}, Astron. J.  \textbf{116,} 1009 (1998).
\bibitem{perlmutter} S. Perlmutter et al, \textit{Measurements of omega and lambda from 42 high-redshift supernovae}, Astron. J.  \textbf{517,} 556 (1999).
\bibitem{Lusso} E. Lusso, E. Piedipalumbo, G. Risaliti, M. Paolillo, S. Bisogni, E. M. A. N. U. E. L. E. Nardini, L. O. R. E. N. Z. O. Amati, \textit{Tension with the flat $\Lambda$CDM model from a high-redshift Hubble diagram of supernovae, quasars, and gamma-ray bursts}, Astron. Astrophys. \textbf{628,} L4(2019).
\bibitem{divalentino H0} Di Valentino E., et al., 2021, Astropart. Phys., 131, 102605.

\bibitem{Riess2019} Riess A. G., Casertano S., Yuan W., Macri L. M., Scolnic D., 2019, The
Astrophysical Journal, 876, 85.

\bibitem{Riess2021}Riess A. G., Casertano S.,YuanW., Bowers J. B., Macri L., Zinn J. C., Scolnic
D., 2021, Astrophys. J. Lett., 908, L6.

\bibitem{Bousso} R. Bousso,  \textit{A covariant entropy conjecture}, J. High Energy Phys. \textbf{1999,} 004 (1999).

\bibitem{hooft} G. 't Hooft, \textit{Dimensional Reduction in Quantum Gravity}, arXiv:gr-qc/9310026.

\bibitem{susskind} L. Susskind, \textit{The world as a hologram}, J. Math. Phys. \textbf{36,} 6377 (1995).

\bibitem{Cohen} A. G. Cohen, D. B. Kaplan, A. E. Nelson   \textit{Effective field theory, black holes, and the cosmological constant}, Phys. Rev. Lett. \textbf{82,} 4971 (1999).

\bibitem{Wang} S. Wang, Y.  Wang,  M. Li, \textit{Holographic dark energy}, Phys. Rep. \textbf{696,} 1 (2017).

\bibitem{Hsu} S. D. Hsu, \textit{Entropy bounds and dark energy}, Phys. Lett. B \textbf{594,} 13 (2004).

\bibitem{Li} M. Li, \textit{A model of holographic dark energy}, Phys. Lett. B \textbf{603,} 1  (2004).

\bibitem{lireview} S. Wang, Y. Wang, M. Li, \textit{Holographic dark energy}, Phys. Rep. 696, 1 (2017).

\bibitem{Granda1} L. N. Granda,  A. Oliveros,  \textit{Infrared cut-off proposal for the holographic density},  Phys. Lett. B  \textbf{669,} 275 (2008).
\bibitem{Oliveros} L. N. Granda, A. Oliveros,  \textit{New infrared cut-off for the holographic scalar fields models of dark energy}, Phys. Lett. B   \textbf{671,} 199 (2009).

\bibitem{Barrow} J. D. Barrow, \textit{The area of a rough black hole}, Phys. Lett. B \textbf{808,} 135643  (2020).

\bibitem{Saridakis} E. N. Saridakis, \textit{Barrow holographic dark energy}, Phys. Rev. D \textbf{102,} 123525  (2020).

\bibitem{Jacobson}T. Jacobson,  \textit{Thermodynamics of spacetime: the Einstein equation of state}, Phys. Rev. Lett. \textbf{75,} 1260  (1995).

\bibitem{Padmanabhan1}T. Padmanabhan,   \textit{Gravity and the thermodynamics of horizons},    Phys. Rep. \textbf{406,} 49 (2005).

\bibitem{Padmanabhan2} T. Padmanabhan,\textit{Thermodynamical aspects of gravity: new insights}, Rep. Prog. Phys. \textbf{73,} 046901 (2010).

\bibitem{Paranjape}A. Paranjape,  S. Sarkar, T. Padmanabhan,  \textit{Thermodynamic route to Field equations in Lanczos-Lovelock Gravity}, Phys. Rev. D  \textbf{74,} 104015 (2006).

\bibitem{Frolov}A. V. Frolov,  L. Kofman,  \textit{Inflation and de Sitter thermodynamics}, J. Cosmol. Astropart. Phys. \textbf{2003,} 009 (2003).

\bibitem{Wang1} B. Wang,  E. Abdalla, R. K. Su, \textit{Relating Friedmann equation to Cardy formula in universes with cosmological constant}, Phys. Lett. B \textbf{503,} 394  (2001).
\bibitem{CaiKim} R. G. Cai and S. P. Kim,
\textit{First Law of Thermodynamics and Friedmann Equations of
Friedmann-Robertson-Walker Universe}, JHEP {\bf0502}, 050 (2005).


\bibitem{Verlinde} E. Verlinde, \textit{On the Origin of Gravity and the Laws of Newton}, J. High Energy Phys. \textbf{1104,} 029 (2011).

\bibitem{Cai} R. G. Cai, L. M. Cao,  N. Ohta, \textit{Friedmann equations from entropic force}, Phys. Rev. D \textbf{81,} 061501 (2010).

\bibitem{Cai1} N. Cai,  L. M. Cao,  Y. P. Hu, \textit{Corrected entropy-area relation and modified Friedmann equations},J. High Energy Phys. \textbf{2008,} 090  (2008).

\bibitem{Sheykhi} A. Sheykhi,   \textit{Entropic corrections to Friedmann equations},  Phys. Rev. D \textbf{81,} 104011 (2010).

\bibitem{Sheykhi1} A. Sheykhi, \textit{Thermodynamics of apparent horizon and modified Friedmann equations}, Eur. Phys. J. C \textbf{69,} 265 (2010).

\bibitem{Sheykhi2} A. Sheykhi, S. H. Hendi,  \textit{Power-law entropic corrections to Newton's law and Friedmann equations},  Phys. Rev. D \textbf{84,} 044023  (2011).

\bibitem{Nojiri} S. I. Nojiri, S. D. Odintsov, E. N.  Saridakis,\textit{Modified cosmology from extended entropy with varying exponent},Eur. Phys. J. C \textbf{79,} 242  (2019).

\bibitem{Abreu} E. M. Abreu, J. A. Neto, E. M.  Barboza Jr, A. C. Mendes,  B. B. Soares, \textit{On the equipartition theorem and black holes non-Gaussian entropies}, Mod. Phys. Lett. A \textbf{35,} 2050266  (2020).

\bibitem{Cai2}  R. G., Cai, S. P. Kim,  \textit{First law of thermodynamics and Friedmann equations of Friedmann-Robertson-Walker universe}, J. High Energy Phys. \textbf{2005,} 050 (2005).

\bibitem{Akbar}  M., Akbar,  R. G. Cai, \textit{Thermodynamic behavior of the Friedmann equation at the apparent horizon of the FRW universe}, Phys. Rev. D \textbf{75,} 084003 (2007).

\bibitem{Sheykhi3} A. Sheykhi, B. Wang,  R.G. Cai, \textit{Thermodynamical Properties of Apparent Horizon in Warped DGP Braneworld}, Nucl. Instrum. Methods Phys. Res. \textbf{779,} 1 (2007).

\bibitem{Sheykhi4} A. Sheykhi,  B.  Wang, R. G. Cai,  \textit{Deep connection between thermodynamics and gravity in Gauss-Bonnet braneworlds}, Phys. Rev. D \textbf{76,} 023515 (2007).

\bibitem{Saridakis1} E. N. Saridakis, \textit{Modified cosmology through spacetime thermodynamics and Barrow horizon entropy}, J. Cosmol. Astropart. Phys. \textbf{07,} 031 (2020).

\bibitem{Sheykhi5}  A. Sheykhi, \textit{Barrow entropy corrections to Friedmann equations}, Phys. Rev. D \textbf{103,} 123503  (2021).

\bibitem{Sheykhi6} A. Sheykhi,   \textit{Modified cosmology through Barrow entropy}, Phys. Rev. D  \textbf{107,} 023505 (2023).

\bibitem{Saridakis:2020zol}E. N. Saridakis, \textit{Barrow holographic dark energy}, Phys. Rev. D \textbf{102,} 123525 (2020).

\bibitem{Srivastava:2020cyk} S. Srivastava, U. K. Sharma, \textit{Barrow holographic dark energy with Hubble horizon as IR cutoff},
 Int. J. Geom. Meth. Mod. Phys. \textbf{18,} 2150014 (2021).

\bibitem{Adhikary:2021xym} P. Adhikary, S. Das, S. Basilakos, E. N. Saridakis, \textit{Barrow holographic dark energy in a nonflat universe}, Phys. Rev. D \textbf{104},  123519 (2021).


\bibitem{Anagnostopoulos:2020ctz} F. K. Anagnostopoulos, S. Basilakos, E. N. Saridakis, \textit{Observational constraints on Barrow holographic dark energy}, Eur. Phys. J. C \textbf{80,} 826 (2020).

\bibitem{Dabrowski:2020atl} M. P. Dabrowski, V. Salzano, \textit{Geometrical observational bounds on a fractal horizon holographic dark energy}, Phys. Rev. D \textbf{102,}  064047 (2020).

\bibitem{Oliveros:2022biu} A. Oliveros, M. A. Sabogal, M. A. Acero, \textit{Barrow holographic dark energy with Granda-Oliveros cutoff}, Eur. Phys. J. Plus \textbf{137}, 783 (2022).

\bibitem{SheBHDE} A. Sheykhi, M. Sahebi Hamedan, \textit{Holographic Dark Energy in Modified Barrow Cosmology}, Entropy {\bf25}, 569 (2023).

\bibitem{SheBADE}A. Sheykhi, S. Ghaffari, \textit{Note on agegraphic dark energy inspired by modified Barrow
entropy}, Physics of the Dark Universe {\bf41}, 101241 (2023).

\bibitem{Sheykhi:2022gzb} A. Sheykhi, B. Farsi, \textit{Growth of perturbations in Tsallis and Barrow cosmology}, Eur. Phys. J. C \textbf{82,}  1111 (2022).


\bibitem{Sheyem}  A. Sheykhi, \textit{Friedmann equations from emergence of cosmic
space}, Phys. Rev. D {\bf87}, 061501(R) (2013).


\bibitem{hayward} S. A. Hayward, \textit{Unified first law of black hole dynamics and relativistic
thermodynamics}, Class. Quant. Grav. {\bf15}, 3147 (1998).


\bibitem{Manoharan:2024thb}
M.~T.~Manoharan,
Eur. Phys. J. C \textbf{84}, no.5, 552 (2024)
doi:10.1140/epjc/s10052-024-12926-z

\bibitem{wetter} C. Wetterich, Nucl. Phys B. 302, 668 (1988).

\bibitem{interact1} Bertolami O, Gil Pedro F and Le Delliou M 2007 Phys. Lett. B
654 165.

\bibitem{oli} G. Olivares, F. Atrio, D. Pavon, Phys. Rev. D 71 (2005) 063523.


\bibitem{Pereira} S. H. Pereira, J. F. Jesus,  \textit{Can dark matter decay in dark energy?}, Phys. Rev. D \textbf{79,} 043517  (2009).

\bibitem{Pavon} D. Pavon,  W. Zimdahl,  \textit{ Holographic dark energy and cosmic coincidence}, Phys. Lett. B \textbf{628,} 206 (2005).


\bibitem{Bennett}C. L. Bennett, M. Bay, M. Halpern,  G. Hinshaw, C. Jackson,  N. Jarosik,  E. L. Wright, \textit{The microwave anisotropy probe* mission},  Astrophys. J. \textbf{583,} 1 (2003).

\bibitem{Spergel1} D. N. Spergel,  L. Verde, H. V. Peiris, E. Komatsu, M. R. Nolta, C. L. Bennett, E. L. Wright,  \textit{First-year Wilkinson Microwave Anisotropy Probe (WMAP)* observations: determination of cosmological parameters}, Astrophys. J. \textbf{148,} 175 (2003).

\bibitem{Tegmark} M. Tegmark,  M. A. Strauss, M. R., Blanton,  K. Abazajian, S. Dodelson, H. Sandvik,  D. G. York, \textit{Cosmological parameters from SDSS and WMAP}, Phys. Rev. D  \textbf{69,} 103501 (2004).

\bibitem{Seljak} U. Seljak,  A. Slosar, P. McDonald,  \textit{Cosmological parameters from combining the Lyman-$\alpha$ forest with CMB, galaxy clustering and SN constraints}, J. Cosmol. Astropart. Phys. \textbf{2006,} 014  (2006).

\bibitem{Spergel2} D. N. Spergel, R. Bean, O. Doré, M. R. Nolta, C. L. Bennett,  J. Dunkley, E. L. Wright,   \textit{Three-year Wilkinson Microwave Anisotropy Probe (WMAP) observations: implications for cosmology}, Astrophys. J. \textbf{170,} 377 (2007).

\bibitem{komatsu} E. Komatsu, et al., Seven-Year Wilkinson Microwave Anisotropy
Probe (WMAP) Observations: Cosmological Interpretation, Astrophys.
J. Suppl. 192 (2011) 18. arXiv:1001.4538, doi:10.1088/0067-
0049/192/2/18.

\bibitem{Sahni} V. Sahni, T. D. Saini, A. A. Starobinsky, U. Alam, \textit{ Statefinder—a new geometrical diagnostic of dark energy},     J. Exp. Theor. Phys. \textbf{77,} 201 (2003).

\bibitem{Alam} U. Alam,  V. Sahni,  T. Deep Saini, A. A. Starobinsky, \textit{Exploring the expanding universe and dark energy using the Statefinder diagnostic}, Mon. Not. R. Astron. Soc  \textbf{344,} 1057 (2003).


\end{thebibliography}
\end{document}